\documentclass[aps,prl,epsf,twocolumn,showpacs,floatfix,preprintnumbers,amsmath,amssymb]{revtex4}
\usepackage{graphicx}

\begin{document} 

\title{Absorbing Phase Transitions with Coupling to a Static Field 
and a Conservation Law}

\author{Julien Kockelkoren and Hugues Chat\'e}

\affiliation{CEA -- Service de Physique de l'\'Etat Condens\'e, Centre d'\'Etudes de Saclay, 91191 Gif-sur-Yvette, France}

\date{\today}

\begin{abstract}
The coupling of branching-annihilating random walks to a static field with
a local conservation law is shown to change the scaling properties
of their phase transitions to absorbing states. In particular, we find
that directed-percolation-class transitions give rise 
to a new universality class distinct
from that characterizing the depinning of the so-called linear 
interface model.
\end{abstract}

\maketitle

Among the many works aiming at an understanding of universality in
out-of-equilibrium critical phenomena, those on transitions to
absorbing states play a leading role because these phase transitions have no
equilibrium counterparts and even occur in one-dimensional (1D)
systems \cite{HH-REVIEW,ODOR-REVIEW}.
In the simple and general
case of the reaction and diffusion of identical particles
$A$ without site occupation restriction,
recent numerical progress (awaiting analytical confirmation)
has led to a global picture involving four basic universality classes
\cite{BARW}. Consider reactions of the type
[$mA\to (m+k)A$, $nA\to (n-l)A$] where $m$, $n$, $k$, and $l$ are positive
integers.
Outside the prominent directed percolation (DP) class, whose simplest
representatives are given by single-particle reactions ($m,n=1$),
the PCPD and TCPD classes 
(for ``pair/triplet contact process with diffusion'' 
---a vocable largely used for historical reasons) are respectively 
characterized by reactions involving two ($m,n=2$) and
three ($m,n=3$) particles \cite{ODOR-BJP}.
The fourth class has been considered to be defined by the conservation
of the parity of the number of particles, hence its usual name
PC (for ``parity-conserving'') \cite{HH-REVIEW}, 
but it is now clear that this is not its defining feature \cite{NOTE}. 
Nevertheless, the relevance of conservation laws is ascertained at 
equilibrium, and it remains important to explore it
within absorbing phase transitions (APT), since conserved quantities abound 
in physical situations.

As a matter of fact, it was argued that DP-class problems
where the order parameter is coupled to an auxiliary (diffusive)
field with conservation
generally show non-DP critical properties \cite{WOH}. The particular case
where the auxiliary field is static was conjectured recently to 
lead to yet another class of APT \cite{ROSSI,PSV2000}.
(Note that the corresponding models
then possess infinitely-many non-connected absorbing states and strong
memory effects.) Numerical simulations confirmed only partially the
above ideas. The situation, in our view, remains 
unsatisfactory for the static case: while in 1D no definite conclusions
could be reached \cite{FES}, 
in 2D and 3D the numerically-estimated critical exponents,
when self-consistent, were found to roughly coincide with those of the 
depinning transition of the so-called linear interface model (LIM)
\cite{LUEBECK}.
A significant departure from DP-scaling was thus found, but
the existence of a separate universality class remains in question.
Meanwhile, no rigorous result is available, although a heuristic
mapping on the dynamics of LIM was advocated in \cite{AM2002}.

In this Letter, we investigate the relevance of conservation laws 
to APT within the more general framework of four ``basic'' universality
classes briefly recalled above, but restricting ourselves to the coupling
to a static auxiliary field:
now the creation of active, diffusive
particles $A$ is conditioned to the presence of passive, static particles
$B$ in such a way that the total number of particles is locally conserved.
We thus study the reactions 
[$mA+kB\to (m+k)A$, $nA\to (n-l)A +lB$], and find evidence for the 
existence of new universality classes. 
We revisit some of the models studied in \cite{ROSSI,PSV2000,FES}, 
but we mostly deal with two-species reaction-diffusion models
derived from those introduced in \cite{BARW}. We show the existence
of a specific ``C-DP'' (``DP with conservation'') class
distinct from both DP and LIM in space dimensions 1 to 3, 
and point at a crucial difference with the LIM which was overlooked in
\cite{AM2002}.
We also show that a specific C-PCPD class exists, 
while no significant departure
from TCPD and PC scaling could be observed in the presence of the coupling
to a static field with conservation.

Extending the notation introduced in \cite{BARW}, we encode the 
rules by the order of the branching and annihilation
reactions using letters {\sf s}, {\sf p}, {\sf t} (standing for
singleton, pair, triplet) for $m,n=1,2,3$, followed by integers $k$ and $l$.
For instance, reactions [$2A+B\to 3A$, $2A\to 2B$] 
are coded as ``conserved PCPD
rule'' {\sf C-pp}12.
Our two-species bosonic models are 
updated in two parallel sub-steps: $A$ particles move to one of their
nearest neighbors (strong diffusion), then on-site reactions take place,
involving the $n_A$ and $n_B$ particles present locally.

We first present our results on the implementation of the
{\sf C-ss}11 rule studied in \cite{PSV2000}. In this case,
the branching reaction $A+B\to 2A$ is performed for each of the 
$n_A$ particles present with probability $p(n_B)= 1-1/2^{n_B}$,
while the annihilation reaction $A\to B$ is performed every $A$ particle
with probability $q$. Note that since $p(0)=0$,
branching is indeed conditioned to the presence of $B$ particles.
Unlike in \cite{PSV2000}, where the total density
of particles is used as a control parameter, here we vary $q$,
which allows to start from perfectly homogeneous initial conditions
(typically one $A$ particle on each site). At large $q$,
annihilation dominates and a static configuration of $B$ particles
is quickly reached, while at small $q$ frequent branching ensures
a stationary density of $A$ particles.

Our numerical methodology is standard: monitoring various order parameters 
(e.g. $\rho_A$ the density of $A$ particles), 
we first determine the critical point $q_{\rm c}$ separating decay to an
absorbing state from sustained activity,
expecting then an algebraic law ($\rho_A\sim t^{-\delta}$)
with $\delta=\beta/\nu_{\parallel}$. 
We record also, during these runs,
the mean squared local gradient $\langle(\nabla h)^2\rangle$
of the interface $h(x,t)$, where $h(x,t)$ is the time integral of the 
local activity at site $x$ until time $t$
($h\to h+1$ whenever $n_A>0$, with $h=0$ initially).
First introduced in \cite{DM2000} for DP-class models,
this fictitious interface was shown to have ``anomalous
scaling'' in the form of local gradients diverging algebraically at
the critical point: $\langle(\nabla h)^2\rangle \sim t^{2\kappa}$. 
(Note that it is also the interface conjectured to behave like in the 
LIM by Alava and Mu\~noz for C-DP critical points \cite{AM2002}.)
After this first series of runs, we estimate the decay of the
(stationary) order parameter with the distance to the previously-estimated
threshold, yielding an estimate of exponent $\beta$. Finally,
exponent $z$ is estimated via the finite-size scaling
of the lifetime of activity at threshold.

\begin{table}
\caption{\label{tab-c-dp} Critical exponents of the  C-DP class
in 1D, 2D, and 3D. The corresponding values for DP and LIM are also given
for reference.
Values for the LIM class, as well as estimates of $\kappa$ for usual
DP models
 were measured by ourselves with system sizes
and timescales of the same order as those used for the C-DP class.
}
\begin{ruledtabular}
\begin{tabular}{clllll}
$d$ &  &    $\delta$     & $z$           & $\beta$       & $2\kappa$ \\ \hline
  &{\sc DP} &  0.1596      & 1.58          & 0.2765        & 0.84(1)    \\
1 &{\sc C-DP} &{\bf 0.140(5)}& {\bf 1.55(3)}  & {\bf 0.29(2)} & {\bf 0.86(1)} \\
  &{\sc LIM}  &  0.125(5)    & 1.43(1)       & 0.25(2)       & 0.35(1)\\ \hline
    &{\sc DP} &    0.451     & 1.76          &  0.584        & 0.56(2)    \\
2 &{\sc C-DP} &{\bf 0.51(1)}  & {\bf 1.55(3)}  & {\bf 0.64(2)}  & 0.50(2)    \\
  &{\sc LIM}  &  0.50(1)      & 1.55(2)        & 0.63(2)        & $0^+$ \\ \hline
    &{\sc DP} &  0.73        & 1.90          & 0.81          & 0.30(5)   \\
3 &{\sc C-DP} &{\bf 0.88(2)} & {\bf 1.73(5)}   & {\bf 0.88(2)}  & $<0.2$    \\
  &{\sc LIM}  &  0.77(2)     & 1.78(7)         & 0.85(2)       & 0    \\
\end{tabular}
\end{ruledtabular}
\end{table}

\begin{figure}
\includegraphics[width=7.8cm,clip]{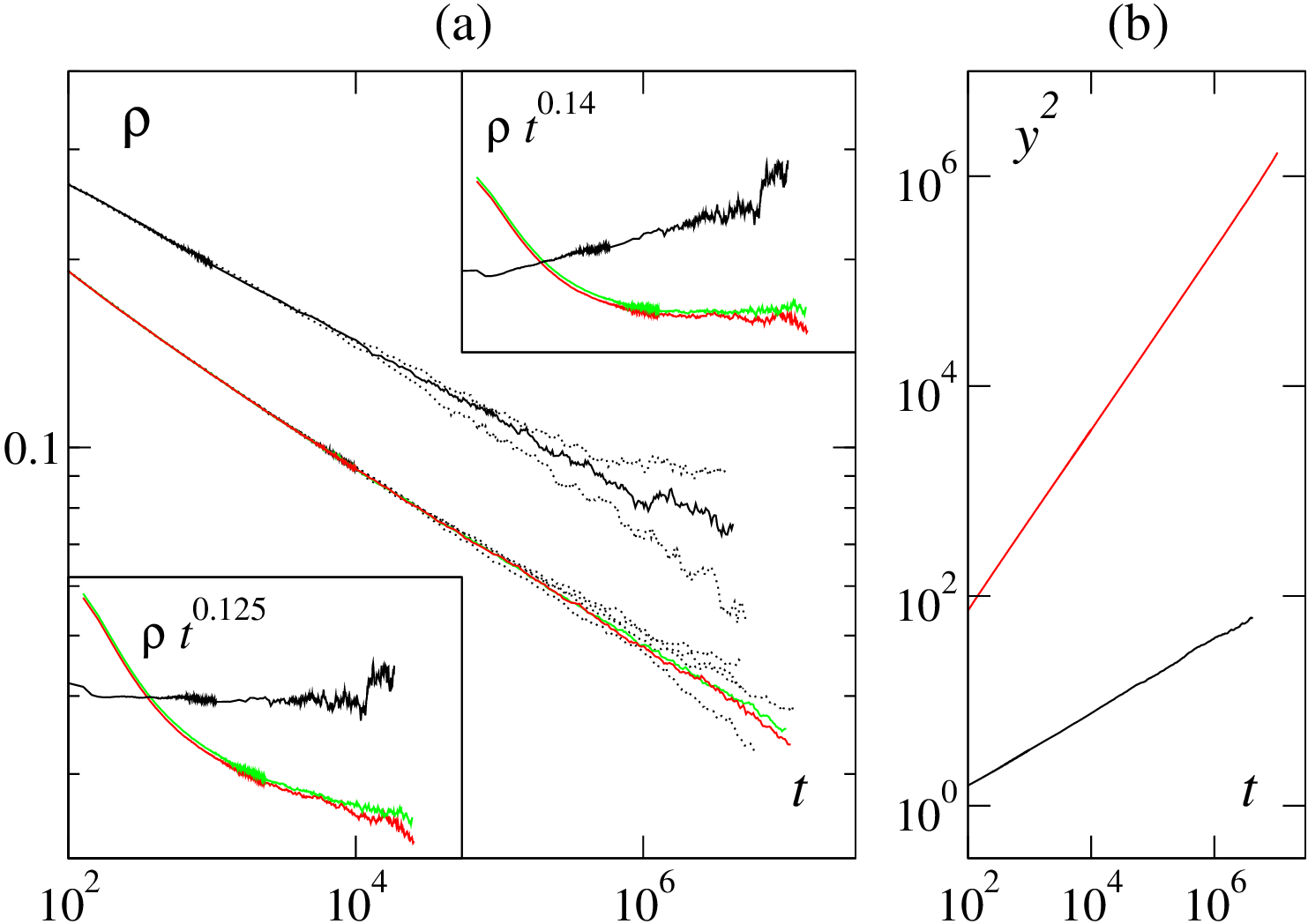}
\caption{1D data from the {\sf C-ss}11 rule implemented as 
in \protect\cite{PSV2000} and from the Leschhorn automaton for the LIM
\protect\cite{LESCHHORN}.
Near-critical decay from fully active homogeneous initial conditions
(system size $2^{22-26}$ sites).
(a): density of active particles $\rho$ vs time (top 3 curves: Leschhorn
model with $p=0.80085, 0.8008, 0.80075$; bottom 5 curves: {\sf C-ss}11
model with $1-q=0.82858, 0.8286, 0.82861, 0.82863, 0.82865$).
Top inset: $\rho\times t^{0.14}$ vs time. 
Bottom inset: $\rho\times t^{0.125}$ vs time.
(b): growth of $y^2=\langle(\nabla h)^2\rangle$
during the same runs at criticality for the {\sf C-ss}11 model
($1-q_{\rm c}=0.82861$, top curve) and for the Leschhorn automaton
($p_{\rm c}=0.8008$, bottom curve).}
\label{f1}
\end{figure}

\begin{figure}
\includegraphics[width=7.8cm,clip]{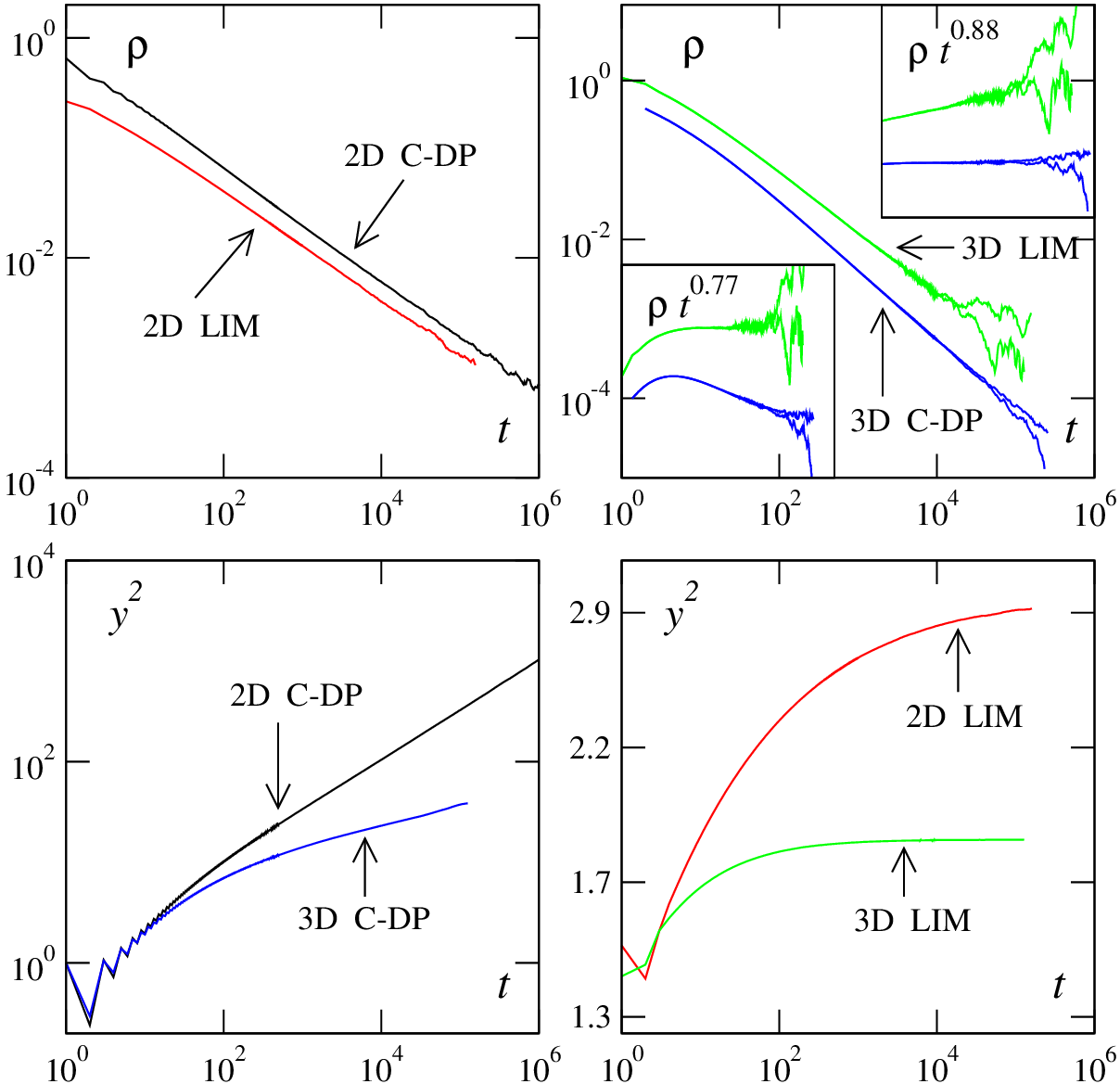}
\caption{Same as Fig.~\ref{f1}, but in 2D and 3D
Top left: near-critical decay of $\rho$ in 2D for 
{\sf C-ss}11 rule (top curve, $1-q=0.65298$, $2^{14\times 2}$ sites)
and Leschhorn automaton (bottom curve, $p=0.64173$,  $2^{15\times 2}$ sites);
Top right: same in 3D for 
{\sf C-ss}11 rule (bottom curves, $1-q=0.57908$ and 0.57909, 
$2^{11\times 3}$ sites)
and Leschhorn automaton (top curves, $p=0.52722$ and 0.527225, 
$2^{10\times 3}$ sites);
Bottom panels:  $y^2=\langle(\nabla h)^2\rangle$ for the same runs
(left: {\sf C-ss}11 rule, right:  Leschhorn automaton).}
\label{f2}
\end{figure}

Table~\ref{tab-c-dp} summarizes our findings for the
{\sf C-ss}11 rule defined above in space dimensions 1 to 3.
Since it is impossible, here, to present all the data which led to this
table, we only show the most significant results, while a full report
will appear elsewhere \cite{TBP}. Two conclusions
can be drawn.
Firstly, we confirm that the coupling to the static particles $B$
together with the conservation law does break the DP class. Our estimates
in 2D and 3D are consistent with those published earlier 
\cite{ROSSI,PSV2000,LUEBECK}, while
in 1D (Fig.~\ref{f1}), our data exhibit more than 3 decades
of scaling, leading to our estimate $\delta^{\rm C-DP}_{\rm 1D}=0.14(1)
\ne\delta^{\rm DP}_{\rm 1D}=0.1596$ (runs showing an effective decay
exponent close to the DP value eventually become clearly subcritical).
Secondly, our results also lead to ruling out 
LIM-class scaling: our estimates of
exponents $\delta$, $\beta$ and $z$
for this {\sf C-ss}11 rule  cannot fully rule out
LIM values (Table~\ref{tab-c-dp}): no discrepancy
could be registered in 2D, while in 1D and 3D we record significant
but small differences (insets of Figs.~\ref{f1} and \ref{f2}).
On the other hand, it is clear from our results on the 
``roughness exponent'' $\kappa$
that there are fundamental differences between a LIM interface and
the fictitious interface constructed at the  {\sf C-ss}11 critical
points: in 1D, the LIM interface shows anomalous scaling, 
but $\kappa$ is more than twice smaller than for the {\sf C-ss}11 rule
(Fig.~\ref{f1}b).
In higher dimensions, LIM interfaces are {\it not} locally rough 
($\kappa=0$), i.e. there is surface tension (Fig.~\ref{f2}).
Meanwhile, we observe, to numerical accuracy, $2\kappa=1-\delta\ne 0$ for
the {\sf C-ss}11 critical points, as observed for ordinary APT \cite{KAPPA}.

In order to assess the robustness of the above results,
we also studied a number of other rules whose ``uncoupled'' version have
APT in the DP-class, implemented in the spirit of \cite{BARW}.
Consider for instance rule {\sf C-sp}12, encoding reactions ($A+B\to2A$, $2A\to 2B$). For each of the $\lfloor n_A/2 \rfloor$ pairs of active particles
present locally, either the branching or the annihilating reaction 
takes place with probability $p$ or $q=1-p$. Here $p$ is independent of $n_B$,
but the branching reaction is further subjected to the presence of 
enough $B$ particles among those present initially.
Finally, if $n_A=1$, only the branching reaction can occur (with probability
$p$ and if $n_B>0$).

\begin{figure}
\includegraphics[width=7.8cm,clip]{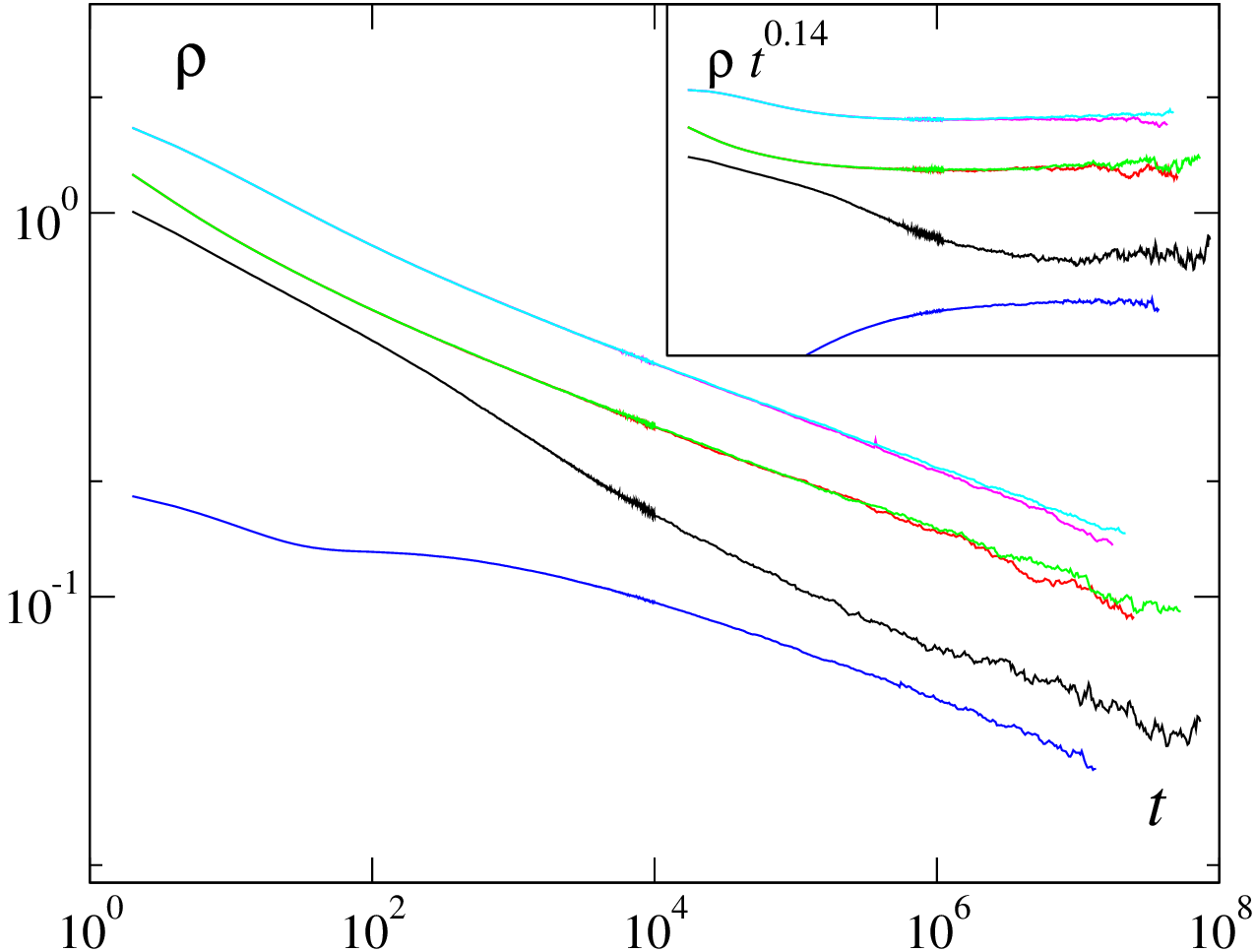}
\caption{Same as Fig.~\protect\ref{f1}a, but for C-DP rules implemented
following \protect\cite{BARW} (see text). Only near-criticality runs are shown.
From top to bottom: 
{\sf C-ss}11 rule ($p=0.720775$, 0.720785, $2^{26}$ sites);
{\sf C-ss}21 rule ($p=0.75905$, 0.75906, $2^{24}$ sites);
{\sf C-sp}12 rule ($p=0.7094$, $2^{22}$ sites);
{\sf C-ps}11 rule ($p=0.83811$, $2^{24}$ sites).
Inset: $\rho \times t^{0.14}$ vs time for the same data.}
\label{f3}
\end{figure}

Rule {\sf C-ss}11, implemented as above,
was investigated in 1, 2, and 3 dimensions, yielding exponent values
equal, to numerical accuracy, to those reported in Table~\ref{tab-c-dp}.
Rules  {\sf C-ss}21,  {\sf C-sp}12, and {\sf C-ps}11 were also studied 
thoroughly in 1D with, again, exponent values fully compatible with those
found for rules {\sf C-ss}11 (Fig.~\ref{f3}). All these results
give credence to the existence of specific C-DP universality class,
distinct from both the DP and the LIM classes.

\begin{figure}
\includegraphics[width=7.8cm,clip]{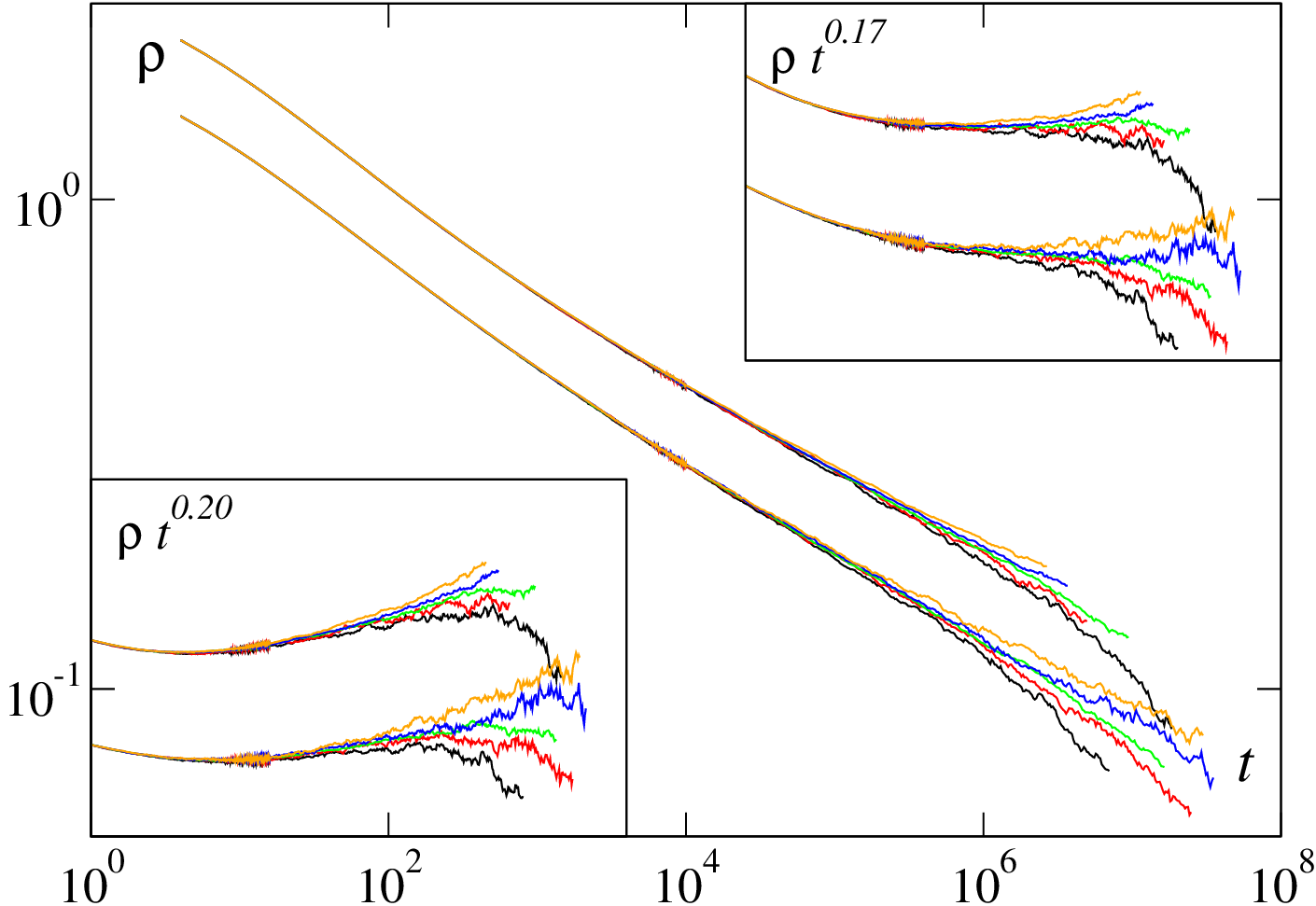}
\caption{Same as Fig.~\protect\ref{f3}, but for C-PCPD rules, for systems
of size $2^{24}$ or $2^{26}$ sites.
Top 5 curves: {\sf C-pp}12 rule (from bottom up: $p=0.81813$, 0.81814,
0.81815, 0.81816, 0.81817).
Bottom 5 curves: {\sf C-pp}22 rule (from bottom up: $p=0.82305$, 0.82308,
0.82309, 0.8231, 0.82312).
Top inset: $\rho\times t^{0.17}$ vs time. 
Bottom inset: $\rho\times t^{0.2}$ vs time.}
\label{f4}
\end{figure}

We now report on our results on conserved PCPD,
TCPD, and PC systems, which we only investigated in 1D.
For rules {\sf C-pp}12 and {\sf C-pp}22, 
our estimates of critical exponents reveal 
an identical departure from the 
PCPD class which we interpret as testifying to the
existence of a C-PCPD class.
In particular, the critical
decay experiments allow to clearly rule out the PCPD value of the 
$\delta$ exponent  (Fig.~\ref{f4}). We find, for both rules studied:
$\delta=0.17(1)$, $\beta=0.32(2)$ and $z=1.55(5)$, 
whereas our estimates for the PCPD class are 
$\delta=0.20(1)$, $\beta=0.37(1)$ and $z=1.70(5)$.
The careful study of rules  {\sf C-tt}12 and  {\sf C-tt}22 did not
reveal any significant departure from TCPD scaling for these C-TCPD
rules. Similarly, we found only small influence of the coupling
to a static field for generalized voter (PC) rules {\sf C-mp}22
and {\sf C-mp}42 \cite{TBP}.
We have, at present, no reason to believe that the C-PC and C-TCPD  
classes ``do not exist''.

Before our main conclusions, 
we come back to the suggested equivalence between the LIM and C-DP
classes which is contradicted by our numerical results. 
In the Leschhorn automaton \cite{LESCHHORN}, used here for our estimates of 
LIM class exponents, an integer-height interface evolves in parallel
according to:
\begin{equation}
\label{lim-eq}
h_i \to h_i+1 \;\;\;\;{\rm iff}\;\;\;\; \nabla^2 h_i + f_i>0
\end{equation}
where $i$ is a lattice index and $\nabla^2h_i$ is the discrete Laplacian
calculated with the nearest-neighbors of $i$. A key ingredient is the quenched
bimodal random ``force'' $f_i$ ($f_i=\pm1$ with probability $p$ and $1-p$)
which is redrawn every time the interface advances. 

The evolution of the fictitious interfaces from which exponent $\kappa$
is estimated in C-DP models can be cast into an equation very similar to
(\ref{lim-eq}) \cite{AM2002}. This is perhaps best seen when considering the
Manna or ``fixed energy'' sandpile, a much-studied model of the C-DP
type which, unfortunately, suffers from severe corrections to
scaling \cite{FES,DICK-XXX,TBP}. There, 
if the number $z_i$ of ``sand grains'' 
present at lattice site $i$ is equal to or exceeds a 
a threshold  $z_{\rm c}$ usually taken to be 2, then
$z_{\rm c}$ grains independently hop to a randomly-chosen nearest-neighbor.
This defines active and passive sites, 
corresponding to the above $A$ and $B$ particles,
and the local conservation law is obviously that of the sand particles.
The fictitious interface thus simply encodes the number of times a site
has discharged its grains since $t=0$: $h_i\to h_i+1$ when $z_i\ge z_{\rm c}$.
This local discharge condition can be expressed in terms of $z_i^{\rm in}$
and  $z_i^{\rm out}$, the cumulated number of particles received from 
and distributed to the neighbors. 
Consider, for simplicity, the 1D case with $z_{\rm c}=2$.
Clearly, $z_i^{\rm out}=2h_i$, whereas
$z_i^{\rm in}$ can be written as the sum of a mean flux 
$(h_{i+1}+h_{i-1})$ and a fluctuating part 
$\tau_i(h_i)$ \cite{AL2001,AM2002}. 
The discharge condition is then expressed 
$$
h_{i+1}+h_{i-1}-2h_i-1+z_i(0)+\tau_i(h_i) >0
$$
where $z_i(0)$ is the initial number of particles on site $i$.
In the general case, the fictitious interface thus obeys
\begin{equation}
\label{fes-eq}
h_i \to h_i+1 \;\;\;{\rm iff}\;\;\; \nabla^2 h_i -1+z_i(0)+\tau_i(h_i) >0
\end{equation}
which is, up to the presumably irrelevant ``columnar noise'' $z_i(0)-1$,
formally equivalent to Eq.(\ref{lim-eq}). There exists, however,
a fundamental difference between the two equations: whereas 
$f_i$ is a delta-correlated noise, $\tau_i$ is a sum of random
variables and its amplitude diverges as $t\to\infty$. As a matter of fact,
since $z_i$ is bounded, both $\langle(\nabla^2 h)^2\rangle$ and 
$\langle\tau(h)^2\rangle$ must diverge in the same way.
We have checked numerically that it is the case, and found moreover
that  $\langle(\nabla^2 h)^2\rangle \sim t^{2\kappa}$.
On the other hand, for the Leschhorn automaton interfaces, 
the amplitudes of both the
noise and the Laplacian remain bounded,
indicating that it constitutes a reasonable physical interface model,
even though  $\langle(\nabla h)^2\rangle$ diverges in 1D. 

These results indicate that the connection between C-DP
and LIM via the fictitious interfaces is only formal and bears no reality. 
Thus, the numerical values of exponents $\delta$, $z$ and $\beta$ are
probably {\it coincidentally} very close to each other in both classes,
and C-DP critical
scaling defines a universality class distinct from that of the LIM.

To sum up, we have found evidence that the coupling of usual
reaction-diffusion systems to a static field with a local conservation
law changes the scaling laws of their transitions to absorbing states,
giving rise to bona fide universality classes distinct from 
previously known ones. Naturally, these results call for rigorous
analytical confirmation which will probably only happen
when the classification of simple APT (DP, PCPD, etc.) is itself 
given some solid theoretical basis.

Reaching a global understanding 
of absorbing phase transitions with coupling to an auxiliary field 
requires, from our viewpoint, to study the 
respective importance of the static or diffusive character of this field
and of the presence of a conservation law.
In particular, the sole coupling to a static field (without conservation)
is usually believed to be irrelevant from existing work on simple
fermionic models such as the pair contact process \cite{PCP}. 
This should now
be tested against the general framework put forward here.
Similarly, the case of a diffusive field with conservation should be
extended beyond the DP-class. Both directions are currently being
investigated.

We thank M.A. Mu\~noz and F. van Wijland for many fruitful discussions.


\begin{thebibliography}{99}

\bibitem{HH-REVIEW} H. Hinrichsen, Adv. Phys. {\bf 49}, 815 (2000).

\bibitem{ODOR-REVIEW} G. \'Odor, eprint cond-mat/0205644, and references
therein.

\bibitem{BARW} J. Kockelkoren and H. Chat\'e, Phys. Rev. Lett.
{\bf 90}, 125701 (2003).

\bibitem{ODOR-BJP} For a recent account of an ongoing debate, see
G. \'Odor, eprint cond-mat/0304023, to appear in Braz. J. Phys.

\bibitem{NOTE} Within our framework, the {\it only} rules 
in the ``parity-conserving'' class are
[$A\to (1+k)A$, $2A\to \emptyset$] with $k$ even, i.e. those which can
be mapped onto models with two symmetric absorbing states, suggesting
that the critical behavior is that of the (generalized) voter model
(see I. Dornic, H. Chat\'e, J. Chave, and H. Hinrichsen, 
Phys. Rev. Lett.  {\bf 87}, 045701 (2001)).


\bibitem{WOH} R. Kree, B. Schaub, and B. Schmittmann, 
Phys. Rev. A {\bf 39}, 2214 (1989);
F. van Wijland, K. Oerding, and H.J. Hilhorst,
Physica A {\bf 251}, 179 (1998); 
K. Oerding, F. van Wijland, J.-P. Leroy, and H.J. Hilhorst,
J. Stat. Phys. {\bf 99}, 1365 (2000).

\bibitem{ROSSI} M. Rossi, R. Pastor-Satorras, and A. Vespignani, 
Phys. Rev. Lett. {\bf 85}, 1803 (2000); 
A. Vespignani, R. Dickman, M.A. Mu\~noz, and S. Zapperi, 
Phys. Rev. E {\bf 62}, 4564 (2000).

\bibitem{PSV2000} R. Pastor-Satorras and A. Vespignani, Phys. Rev. E {\bf 62}, R5875 (2000).

\bibitem{FES} R. Dickman et al., Phys. Rev. E {\bf 64}, 056104 (2001).

\bibitem{LUEBECK} S. L\"ubeck, Phys. Rev. E {\bf 64}, 016123 (2001);
{\bf 66}, 046114 (2002).

\bibitem{AM2002} M. Alava and M.A. Mu\~noz, Phys. Rev. E {\bf 65}, 026145 (2002).

\bibitem{DM2000} R. Dickman and M.A. Mu\~noz, Phys. Rev. E {\bf 62}, 7632 (2000).

\bibitem{TBP} J. Kockelkoren and H. Chat\'e, to be published.

\bibitem{KAPPA} J. Kockelkoren et al., in preparation.

\bibitem{LESCHHORN} H. Leschhorn, Physica A {\bf 195}, 324 (1993);
M. Jost and K.D. Usadel, Physica A {\bf 255}, 15 (1998).

\bibitem{DICK-XXX} R. Dickman and J.M.M. Campelo, eprint cond-mat/0301054;
R. Dickman, eprint cond-mat/0209589.

\bibitem{AL2001} M. Paczuski, S. Maslov, and P. Bak, 
Phys. Rev. E {\bf 53}, 414 (1996);
M.J. Alava and K.B. Lauritsen, 
Europhys. Lett. {\bf 53}, 563 (2001).

\bibitem{PCP} I. Jensen and R. Dickman, Phys. Rev. E {\bf 48},1710 (1993);
R. Dickman, Phys. Rev. E {\bf 53}, 2223 (1996);
M.A. Mu\~noz, G. Grinstein, R. Dickman, and R. Livi, 
Phys. Rev. Lett. {\bf 76}, 451 (1996); J. Stat. Phys. {\bf 91}, 541 (1998).


\end{thebibliography}
\end{document}